\documentclass[12pt,a4paper,final]{iopart}
\usepackage{iopams}  
\usepackage{graphicx}
\usepackage[breaklinks=true,colorlinks=true,linkcolor=blue,urlcolor=blue,citecolor=blue]{hyperref}
\usepackage{dcolumn}
\usepackage{rotating} 
\usepackage{changepage} 

\begin{document}

\title[Low coverage surface diffusion in complex energy landscapes]{Low coverage surface diffusion in complex energy landscapes: Analytical solution and application to intercalation in topological insulators}

\author{Miguel~A.~Gosalvez$^{1,2,3}$, Mikhail~M.~Otrokov$^{4,2}$, Nestor~Ferrando$^{3,2}$, Anastasia~G.~Ryabishchenkova$^{4}$, Andres~Ayuela$^{3,2}$, Pedro~M.~Echenique$^{1,2,3}$, Eugene~V.~Chulkov$^{1,2,3,4}$}

\address{$^1$
Dept. of Materials Physics, University of the Basque Country UPV/EHU, 20018 Donostia-San Sebastian, Spain
}
\address{$^2$
Donostia International Physics Center (DIPC), 20018 Donostia-San Sebastian, Spain
}
\address{$^3$
Centro de F\'isica de Materiales CFM-Materials Physics Center MPC, centro mixto CSIC -- UPV/EHU, 20018 Donostia-San Sebastian, Spain
}
\address{$^4$
Tomsk State University, 634050 Tomsk, Russia
}

\ead{miguelangel.gosalvez@ehu.es}

\date{\today}

\begin{abstract}
A general expression is introduced for the tracer diffusivity in complex periodic energy landscapes with more than one distinct hop rate in two- and three-dimensional diluted systems (low coverage, single-tracer limit). For diffusion in two dimensions, a number of formulas are presented for complex combinations of hop rates in systems with triangular, rectangular and square symmetry. The formulas provide values in excellent agreement with Kinetic Monte Carlo simulations, concluding that the diffusion coefficient can be directly determined from the proposed expressions without performing such simulations. Based on the diffusion barriers obtained from first principles calculations and a physically-meaningful estimate of the attempt frequencies, the proposed formulas are used to analyze the diffusion of Cu, Ag and Rb adatoms on the surface and within the van der Waals (vdW) gap of a model topological insulator, Bi$_{2}$Se$_{3}$. Considering the possibility for adsorbate intercalation from the terraces to the vdW gaps at morphological steps, we infer that, at low coverage and room temperature: (i) a majority of the Rb atoms bounce back at the steps and remain on the terraces, (ii) Cu atoms mostly intercalate into the vdW gap, the remaining fraction staying at the steps, and (iii) Ag atoms essentially accumulate at the steps and gradually intercalate into the vdW gap. These conclusions are in good qualitative agreement with previous experiments. \bf{Supplementary Data is provided}.
\end{abstract}

%
%

\vspace{5 mm}

\submitto{\NJP}


\section{Introduction}

The diffusion of atoms and molecules on crystalline surfaces is fundamental to several technologies \cite{Ala-Nissila, Naumovets, Gomer, Kellogg, Ferrando}. This includes heterogeneous catalysis for mass production of essential compounds in the chemical, food and energy industries \cite{Ma_HeterogeneousCatalysisByMetals}, as well as the growth of thin films for the fabrication of semiconductor devices and novel two-dimensional (2D) materials, such as graphene \cite{Li_Science, Bae_NatureNanotechnology}. Planar synthesis technologies, such as Chemical Vapor Deposition (CVD),  where surface diffusion plays a key role, are currently attracting increasing attention as an alternative to supply a complete, new generation of atom-thick materials, including semi-metals (graphene, NiTe$_2$, VSe$_2$,...)\cite{Li_Science,Bae_NatureNanotechnology, Bonaccorso2012564}, semiconductors (WS$_2$, WSe$_2$, MoS$_2$, MoSe$_2$, MoTe$_2$, TaS$_2$, RhTe$_2$, PdTe$_2$,...)\cite{Bonaccorso2012564, Xu_GrapheneLike2DMaterials, Wang_2DelectronicsOnCVDMoS2, Lee_SynthesisOfMoS2byCVD}, insulators (hexagonal-BN, HfS$_2$,...)\cite{Xu_GrapheneLike2DMaterials, Ismach_hBN, Ci_hBN}, superconductors (NbS$_2$, NbSe$_2$, NbTe$_2$, TaSe$_2$,...)\cite{Bonaccorso2012564, Boscher_SynthesisOfNbSe2byAPCVD} and topological insulators (Bi$_2$Se$_3$, Bi$_2$Te$_3$, Sb$_2$Te$_3$)\cite{Yan_SynthesisOfBi2Se3byCVD, Li_SynthesisOfBi2Se3byvdWEpitaxy, Adroguer2012}. Recently, the deposition of various adsorbates on model topological insulators, such as Bi$_{2}$Se$_{3}$ and Sb$_{2}$Te$_{3}$, has received much consideration \cite{Bianchi1, Zhu, Bianchi2, Valla, Wang, Ye, Scholz, Seibel.prb2012, Eremeev.njp2012, Otrokov.jetpl2012, Eelbo2013}. Adsorbate deposition provides a route to control the position of the Dirac point relative to the Fermi level \cite{Valla, Seibel.prb2012}. Structural investigations of the impurity-deposited Bi$_2$Se$_3$ surface reveal partial \cite{Bianchi2} or almost complete \cite{Wang,Ye.arxiv2011} loss of the adatoms at room and higher temperatures, indicating that the adsorbates may diffuse across the terraces and intercalate at the steps into the van der Waals (vdW) gaps \cite{Otrokov.jetpl2012}. In addition to general applications in energy storage and synthesis of atom-thick materials, intercalation offers the possibility of adjusting the properties of the host material, {\it e.g.} converting a topological insulator into a superconductor \cite{Koski_IntercalationOfMetalsIntoBi2Se3, Hor_SuperconductivityByDopingTopologicalInsulator, PhysRevLett104057001, Wray_SuperconductivityByDopingTopologicalInsulator, Diamantini}. In this manner, the understanding of adsorbate diffusion in material-specific energy landscapes remains a prerequisite for the clarification of the novel properties observed in new materials.

\begin{figure}[htb]
\begin{center}
\includegraphics[width=11.0 cm]{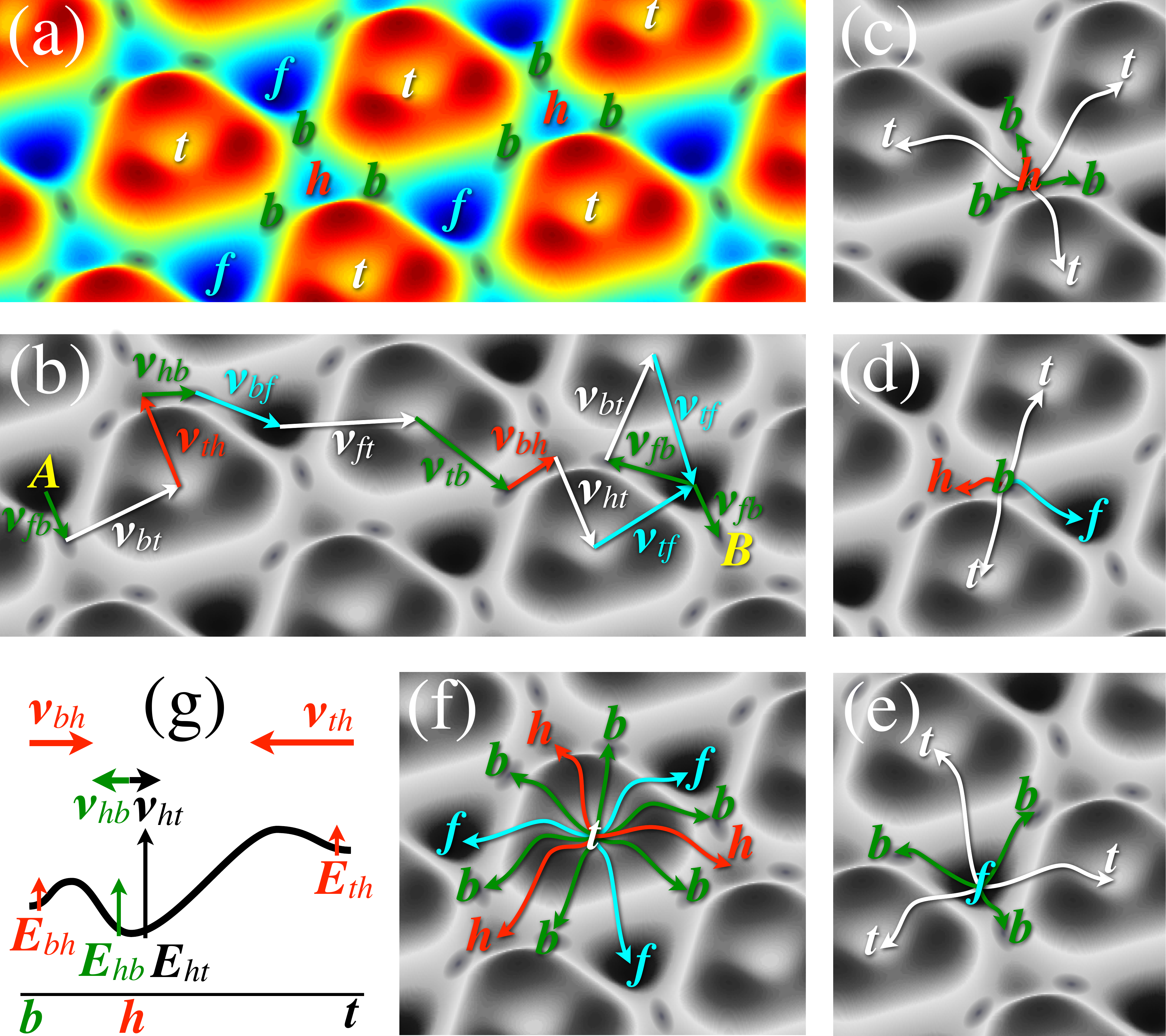}
\end{center}
\caption{
(a) Example of a complex potential energy landscape for a diffusing particle. Four 
adsorption site types are indicated: $f$, $h$, $t$ and $b$. (b) A possible 
diffusion track (random walk) involving 14 performed jumps with 10 different 
hop rates $\nu_{ij}$. (c)-(f) Possible hops from each site type ($h$, $b$, 
$f$ and $t$, respectively). 
Gray-shaded landscapes in (b)-(f) are used to highlight the colored arrows/hops.
(g) Energy paths for hops starting/ending in $h$ 
sites. A longer arrow assigned to a hop rate ($\nu_{th}$, $\nu_{ht}$, etc...) 
indicates a larger rate.
}
\label{Figure_ComplexEnergyLandscape}
\end{figure}

In this study we are interested in the surface diffusion of an adparticle 
on a complex periodic energy landscape, such as the one shown in 
figure \ref{Figure_ComplexEnergyLandscape}(a). One can discern the presence of 4 
different adsorption sites (labeled as $f$, $h$, $t$ and $b$), 
corresponding to the locations where the energy has a local/global minimum. 
A diffusing particle proceeds by hopping between the sites, as 
shown schematically in figure \ref{Figure_ComplexEnergyLandscape}(b). If one 
focuses on a particular site, as illustrated in  
figure \ref{Figure_ComplexEnergyLandscape}(c)-(f) for $h$, $b$, $f$ and $t$, 
respectively, one may notice that each hop requires surpassing a different energy barrier.
As further emphasized in  figure \ref{Figure_ComplexEnergyLandscape}(g), 
the different energy barriers and 
dissimilar shapes of the energy wells (= dissimilar attempt frequencies) 
result in different hop rates for the different jumps ($\nu_{ht}$, $\nu_{th}$, 
$\nu_{hb}$, $\nu_{bh}$, {\it etc...}), including the forward and backward 
directions. As a result, the random walk between points $A$ and 
$B$ in  figure \ref{Figure_ComplexEnergyLandscape}(b) involves as many as 
10 different hop rates for a total of 14 performed hops. 
In this study we focus on describing analytically the average distance travelled by the adparticle as a function of the hop rates $\nu_{ij}$ when the number of hops grows very large, {\it i.e.} the diffusion time becomes arbitrarily large.

The average squared distance covered by a single particle per unit time is 
a well-defined quantity, known as the {\it tracer diffusion coefficient} 
(or {\it tracer diffusivity}) \cite{Ala-Nissila}:
\begin{equation}
D_{T} =  \frac{1}{2\alpha} \lim_{t \rightarrow \infty} \frac{ \sum_{i=1}^{n} \left< |{\bf r}_{i}(t) - {\bf r}_{i}(0)|^{2} \right> }{nt} ,
\end{equation}
where $\alpha=1,2,3$ is the number of dimensions, $n$ is the number 
of adparticles simultaneously present on the surface, ${\bf r}_{i}(t)$ 
designates the position of adparticle $i$ at time $t$, and 
$\left< \cdot \right>$ is the ensemble average. Not surprisingly, $D_{T}$ 
is a function of the number of adparticles $n$ or, equivalently, of the 
coverage $\theta=n/m$, where $m$ is the number of adsorption sites that 
may be occupied by the adparticles. The larger the number of adparticles the 
smaller the number of available empty sites where any chosen adparticle 
can jump to, thus leading to correlation effects between consecutive hops, 
also known as memory effects \cite{Ala-Nissila, Naumovets}.
This is specially relevant for systems 
with strong adsorbate-adsorbate interactions \cite{Vattulainen1999, 
Vattulainen1998}.

We are interested in the low coverage regime, where the adparticle density is so low that the chance of affecting each other's motion is negligible:
\begin{equation}
D^{\theta \approx 0}_{T} =  \frac{1}{2\alpha} \lim_{t \rightarrow \infty} \frac{ \left< |{\bf r}(t) - {\bf r}(0)|^{2} \right> }{t} .
\label{Eq_D_T}
\end{equation}
The low coverage limit is an important 
measure as it provides a simple procedure to compare the typical distances 
covered by different adsorbates across different substrates \cite{Ala-Nissila, 
Naumovets, Gomer, Kellogg, Ferrando}. Previous analytical work on diluted 
systems has focused on the determination of the center-of-mass diffusivity 
for small 2D islands and clusters on metal surfaces based on the 
master-equation \cite{Salo2001, SanchezEvans1999, TitulaerDeutsch1982} or 
the continuous time random walk formalism \cite{Kley1997, Haus1987}.
In the  framework of bulk-mediated surface diffusion, Revelli {\it et al.} described the average motion of the adsorbed molecules, including both Markovian and non-Markovian desorption, by using the generalized master-equation approach \cite{Revelli2005}.
Birnie {\it et al.} \cite{Birnie1990}
and Condit {\it et al.} \cite{Condit1969} derived the overall jump rate for complex, sequential diffusion paths in three-dimensional (3D) crystals, where typical vacancy-interstitial 
complexes evolve by repeating a particular sequence of hops.
The present study generalizes such sequential analysis 
by providing a universal expression for the diffusivity in complex hopping networks where both parallel 
and sequential diffusion routes are available between 
the different adsorption sites.

Computationally, adsorbate diffusion is traditionally studied by \cite{Ala-Nissila, Ferrando}: 
(i) first principles calculations, typically involving the use of density functional theory (DFT) 
for the determination of the activation barriers and attempt frequencies,
which are then passed to other methods; 
(ii) molecular dynamics simulations, which numerically solve Newton's 
equations for the substrate and adsorbate atoms based on effective interaction 
potentials, enabling the analysis at the picosecond time scale for systems with 
$\sim$10$^5$ atoms; (iii) Langevin models, which describe the adparticle in an 
effective periodic force field,
restricting the analysis to general 
trends for time scales of picoseconds; and (iv) Kinetic Monte Carlo (KMC) 
simulations, which focus on describing the hops between adjacent 
basins ({\it rare events}) while disregarding all other vibrations, 
this way enabling long simulated times (seconds and minutes) with affordable 
computational resources. 
In this study we validate the proposed formulas for the diffusivity by direct comparison to KMC simulations in relevant energy landscapes, finally using the formulas to discuss the relative mobility and intercalation of various adsorbates in the context of topological insulators.

\section{Theory}

\begin{figure}[htb]
\begin{center}
\includegraphics[width=11.0 cm]{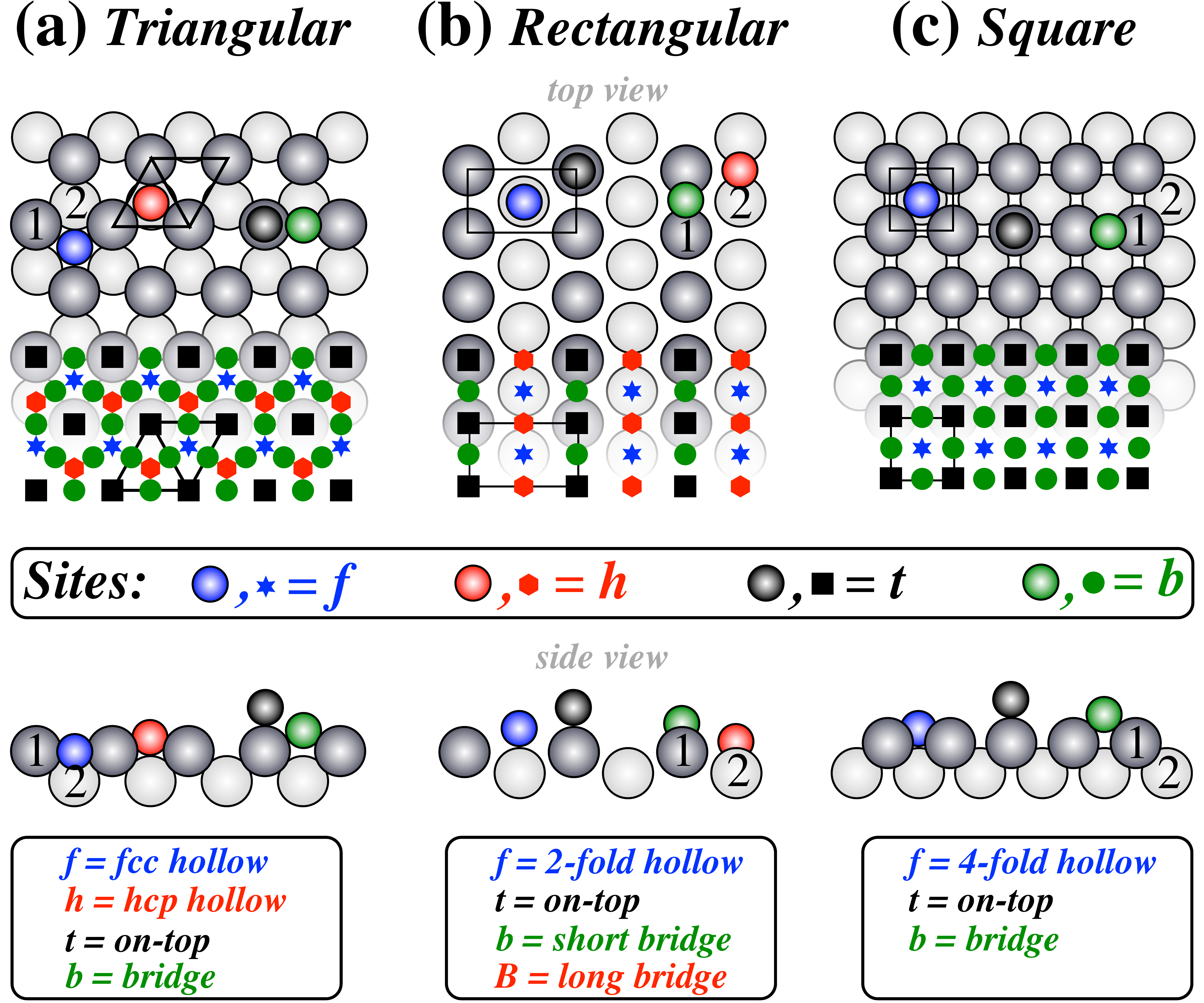}
\end{center}
\caption{
Schematic illustration of three typical surfaces with (a) triangular symmetry [{\it e.g.} fcc(111) and hcp(0001)], (b) rectangular symmetry [{\it e.g.} fcc(110)], and (c) square symmetry [{\it e.g.} fcc(100)]. 
We refer to typical adsorption sites as $f$ (for {\it fcc hollow}), $h$ (for {\it hcp hollow}), $t$ (for {\it on-top}), $b$ (for {\it bridge/short bridge}), and $B$ (for {\it long bridge}).  
}
\label{Figure_TriangularRectangularSquaredSurfaces}
\end{figure}

Let us consider a system with $S$ different site types, such as the one shown in figures \ref{Figure_TriangularRectangularSquaredSurfaces}(a) and \ref{Figure_TriangularRectangularSquaredSurfaces}(b) for $S=4$, or figure \ref{Figure_TriangularRectangularSquaredSurfaces}(c) for $S=3$. 
Although we assume diffusion in two dimensions, the underlying mathematical treatment and main result are also valid for three dimensions.
For a generic hop from site type $i$ to site type $j$ ($i,j=1,...,S$) we consider that the hop distance $l_{ij}$, hop rate $\nu_{ij}$ and hop multiplicity $n_{ij}$ are known, and we define a new variable, the {\em rateplicity} $\mu_{ij}$, as the product of the rate and the multiplicity:
\begin{equation}
\mu_{ij}=n_{ij}\nu_{ij} \;.
\end{equation}
We then propose that the low coverage diffusivity $D^{\theta \approx 0}_{T}$, as defined in equation \ref{Eq_D_T}, can be written as a weighted sum of partial diffusivities:
\begin{equation}
D^{\theta \approx 0}_{T} = \frac{1}{2\alpha} \Sigma_{i} w_{i} (\Sigma_{j} \mu_{ij} l_{ij}^2) \;,  \label{D_5}
\end{equation}
where $(\Sigma_{j} \mu_{ij} l_{ij}^2)$ is the partial diffusivity from site $i$, which contains the rateplicities and hop lengths for all jumps from site type $i$ to any accessible site type $j$, and the dimensionless coefficient $w_{i}$ (or normalized weight for site $i$) stands for the probability to find the adatom at site type $i$: 
\begin{equation}
w_{i} = \frac{ B_{i} } { \Sigma_{j} B_{j} }  \; \; \; (i=1,...,S), \label{Definition_b_i}
\end{equation}
where the $B_{i}$ coefficients consist of sums and products of rateplicities, their particular form depending on the number of adsorption sites $S$. As an example, for $S=2$ and $S=3$ we have:
\begin{eqnarray}
\textstyle B_{i} = & \textstyle \mu_{ji}  \; \; \; & \textstyle (S=2), \label{Definition_B_i_S2} \\
\textstyle B_{i} = & \textstyle \mu_{ji}(\mu_{ki}+\mu_{kj})+\mu_{jk}\mu_{ki}  \; \; \; & \textstyle (S=3), \label{Definition_B_i_S3}
\end{eqnarray}
while a slightly more complex definition can be used for $S=4$:
\begin{eqnarray}
\textstyle B_{i} & \textstyle = & \textstyle B_{ji} = B_{j'i} \; \; \; (i \ne j \ne j') \label{Definition_B_i} \\
\textstyle B_{ij} & \textstyle = & \textstyle A_{ik}^{l}\mu_{kj} + A_{il}^{k}\mu_{lj} + U_{l}^{k}\mu_{ij} \; \; \; (i \ne j \ne k \ne l) \label{Definition_B_ij} \\
\textstyle A_{ij}^{k} & \textstyle = & \textstyle \mu_{ij}(R_{k}-\mu_{kk}) + \mu_{ik}\mu_{kj} \; \; \; (i \ne j \ne k) \label{Definition_A_ij^k} \\
\textstyle U_{j}^{i} & \textstyle =  & \textstyle (R_{i}-\mu_{ii})(R_{j}-\mu_{jj}) - \mu_{ij}\mu_{ji} \; \; \; (i \ne j) \label{Definition_U_j^i} \\
\textstyle & \textstyle = & \textstyle A_{jk}^{i} + A_{jl}^{i} \; \; = \;\; \textstyle A_{ik}^{j} + A_{il}^{j} \;\; = \;\; \textstyle U_{i}^{j}
\end{eqnarray}

In practice, to determine each $B_{i}$ it is convenient to regard $i$ as the end site of a jump from a neighboring site $k$, so that $B_{i}=B_{ki}$, which is then determined by applying equations \ref{Definition_B_ij}-\ref{Definition_U_j^i} ($S=4$) or equation \ref{Definition_B_i_S3} ($S=3$) or equation \ref{Definition_B_i_S2} ($S=2$). Any value of $k$ different from $i$ can be used since the $B_{ij}$ coefficients depend only on the second index ($B_{i}=B_{ki}=B_{k'i}$, see equation \ref{Definition_B_i}). This is easily demonstrated by writing out $B_{ki}$ and $B_{k'i}$ according to equation \ref{Definition_B_ij} and confirming their correspondence. For $S \ge 5$, a general procedure to determine the $B_{i}$ coefficients is described in the {\bf Supplementary Data}, where also a rigorous derivation of equation \ref{D_5} is provided. Valid for any value of $S$ in two and more dimensions, equation \ref{D_5} is our central result.

Although equation \ref{D_5} considers all $S \times S$ hops between all possible pairs from a set of $S$ different site types, non-occurring jumps can be eliminated from equation \ref{D_5} by setting their rateplicities to zero ($\mu_{ij}=0$). This may lead, however, to undetermined values, {\it e.g.} $b_{i} = \frac{0}{0}$, and it is better to substitute the zeroed rateplicities by a small value ($\epsilon$) and take the limit $\epsilon \rightarrow 0$. As an example for the system shown in  figure \ref{Figure_ComplexEnergyLandscape}, if we consider only the $fh$, $ft$, $hf$, $ht$ and $tf$ hops (with multiplicity equal to 3) while disregarding all other hops, we may set $\mu_{ff}=\mu_{fb}=\mu_{hh}=\mu_{hb}=\mu_{th}=\mu_{tt}=\mu_{tb}=\mu_{bf}=\mu_{bh}=\mu_{bt}=\mu_{bb}=\epsilon$ in order to calculate the $B_{i}$ coefficients and then take the limit $\epsilon \rightarrow 0$ to determine the $b_{i}$ factors. We have:
\begin{eqnarray}
\textstyle B_{f} & \textstyle = & \textstyle B_{hf} = B_{tf}  = A_{th}^{b}\mu_{hf} + A_{tb}^{h}\mu_{bf} + U_{b}^{h}\mu_{tf} \\
\textstyle & \textstyle = & \textstyle \epsilon(\mu_{hf}+\mu_{ht})\mu_{tf} + O(\epsilon^2) \\
\textstyle B_{h} & \textstyle = & \textstyle B_{fh} = A_{ft}^{b}\mu_{th} + A_{fb}^{t}\mu_{bh} + U_{t}^{b}\mu_{fh} \\
\textstyle & \textstyle = & \textstyle \epsilon\mu_{tf}\mu_{fh} + O(\epsilon^2) \\
\textstyle B_{t} & \textstyle = & \textstyle B_{ht} = B_{ft} = A_{fh}^{b}\mu_{ht} + A_{fb}^{h}\mu_{bt} + U_{h}^{b}\mu_{ft} \\
\textstyle & \textstyle = & \textstyle \epsilon(\mu_{fh}\mu_{ht}+\mu_{hf}\mu_{ft}+\mu_{ft}\mu_{ht}) + O(\epsilon^2) \\
\textstyle B_{b} & \textstyle = & \textstyle  0
\end{eqnarray}
Thus, using Equations  \ref{D_5} and \ref{Definition_b_i} the diffusivity is: 
\begin{eqnarray}
\textstyle D^{\theta \approx 0}_{T} & \textstyle = & \textstyle \frac{1}{2\alpha} \frac{ \left[ B_{f}(\mu_{fh}+\mu_{ft}) + B_{h} (\mu_{hf}+\mu_{ht}) + B_{t}\mu_{tf} \right] l^2 } { B_{f} + B_{h} + B_{t}} \\
\textstyle & \textstyle = & \textstyle  \frac{3}{2\alpha} \frac{2(\nu_{fh}+\nu_{ft})(\nu_{hf}+\nu_{ht})+\nu_{fh}\nu_{ht}}{\nu_{fh}(1+\frac{\nu_{ht}}{\nu_{tf}}) + (\nu_{hf}+\nu_{ht})(1+\frac{\nu_{ft}}{\nu_{tf}}) } l^2
\end{eqnarray}

\begin{table*}[htbp]
   \caption{
   Examples of low coverage tracer diffusivities ($D^{\theta \approx 0}_{T}$) for different combinations of hop rates between standard adsorption sites on square, rectangular and triangular lattices. Site labels: {\it f = 4-fold hollow (square) / 2-fold hollow (rectangular) / fcc hollow (triangular)}, {\it h = hcp hollow (triangular)}, {\it t = on-top}, {\it b = bridge (square) / short-bridge (rectangular)} and {\it B = long-bridge (rectangular)}. See the {\bf Supplementary Data} for additional formulas.
   }
 \begin{adjustwidth}{-20 mm}{}
 \renewcommand{\tabcolsep}{0.5pt}
   \centering
   \label{Table_ExampleEquations}
   
   \vspace{0.15cm}         

   \begin{tabular}{@{} |c|r|c|c| @{}} 

      \hline

   \mbox{ \footnotesize {\bf Sym} }  & \mbox{ \footnotesize {\bf Rate / multiplicity / distance} } & \mbox{ \footnotesize {\bf Geometry} } & \mbox{ \footnotesize {\bf $D^{\theta \approx 0}_{T}$ [$\alpha=2$]} } \\

      \hline
      \hline
      
      \begin{sideways}\hspace{-18 mm}Square\end{sideways}
      &
\begin{tabular}{ccccc} 
 \mbox{ \footnotesize $\nu_{ff}$ } &  \mbox{ \footnotesize $\nu_{ft}$ } &  \mbox{ \footnotesize $\nu_{tf}$ } &  \mbox{ \footnotesize $\nu_{tt}$ } &  \mbox{ \footnotesize  }\\
 \mbox{ \footnotesize $4$ } &  \mbox{ \footnotesize $4$ } &  \mbox{ \footnotesize $4$ } &  \mbox{ \footnotesize $4$ } &  \mbox{ \footnotesize  }\\
 \mbox{ \footnotesize $l$ } &  \mbox{ \footnotesize $d$ } &  \mbox{ \footnotesize $d$ } &  \mbox{ \footnotesize $l$ } &  \mbox{ \footnotesize }\\
\end{tabular}
      &
\begin{tabular}{c} 
 \vspace{-3 mm}\\
 \includegraphics[height=1.1 cm]{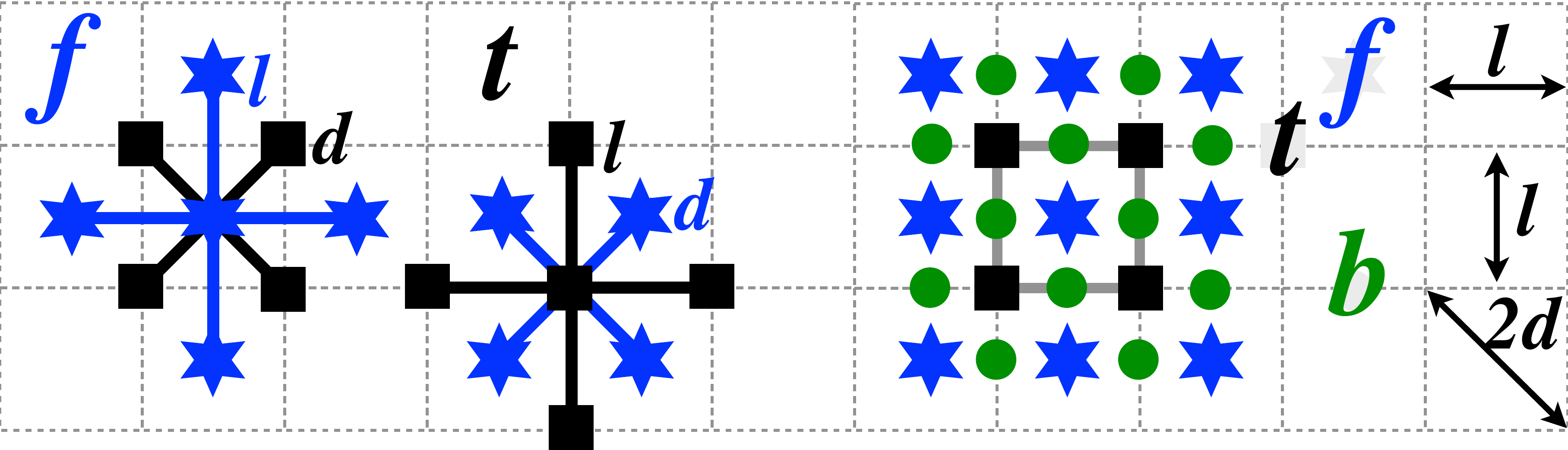}
\end{tabular}
      &
      \mbox{ \footnotesize $ \frac{4}{2\alpha} \frac{2\nu_{ft}\nu_{tf}d^2+(\nu_{tf}\nu_{ff}+\nu_{ft}\nu_{tt})l^2}{\nu_{ft}+\nu_{tf}}$ }
      \\

      \cline{2-4} 
      
      &
\begin{tabular}{r} 
\begin{tabular}{ccccc} 
 \mbox{ \footnotesize $\nu_{ii}$ } &  \mbox{ \footnotesize $\nu_{ib}$ } &  \mbox{ \footnotesize $\nu_{bi}$ } &  \mbox{ \footnotesize $\nu_{bb}^{d}$ } &  \mbox{ \footnotesize $\nu_{bb}^{l}$ }\\
 \mbox{ \footnotesize $4$ } &  \mbox{ \footnotesize $4$ } &  \mbox{ \footnotesize $2$ } &  \mbox{ \footnotesize $4$ } &  \mbox{ \footnotesize $2$ }\\
 \mbox{ \footnotesize $2d$ } &  \mbox{ \footnotesize $l/2$ } &  \mbox{ \footnotesize $l/2$ } &  \mbox{ \footnotesize $d$ } &  \mbox{ \footnotesize $l$ }\\
\end{tabular}
\\
\mbox{ \footnotesize $i=f$ or $i=t$ }
\\
\end{tabular}
      &
\begin{tabular}{ccc} 
  \vspace{-3 mm}\\
  \includegraphics[height=1.1 cm]{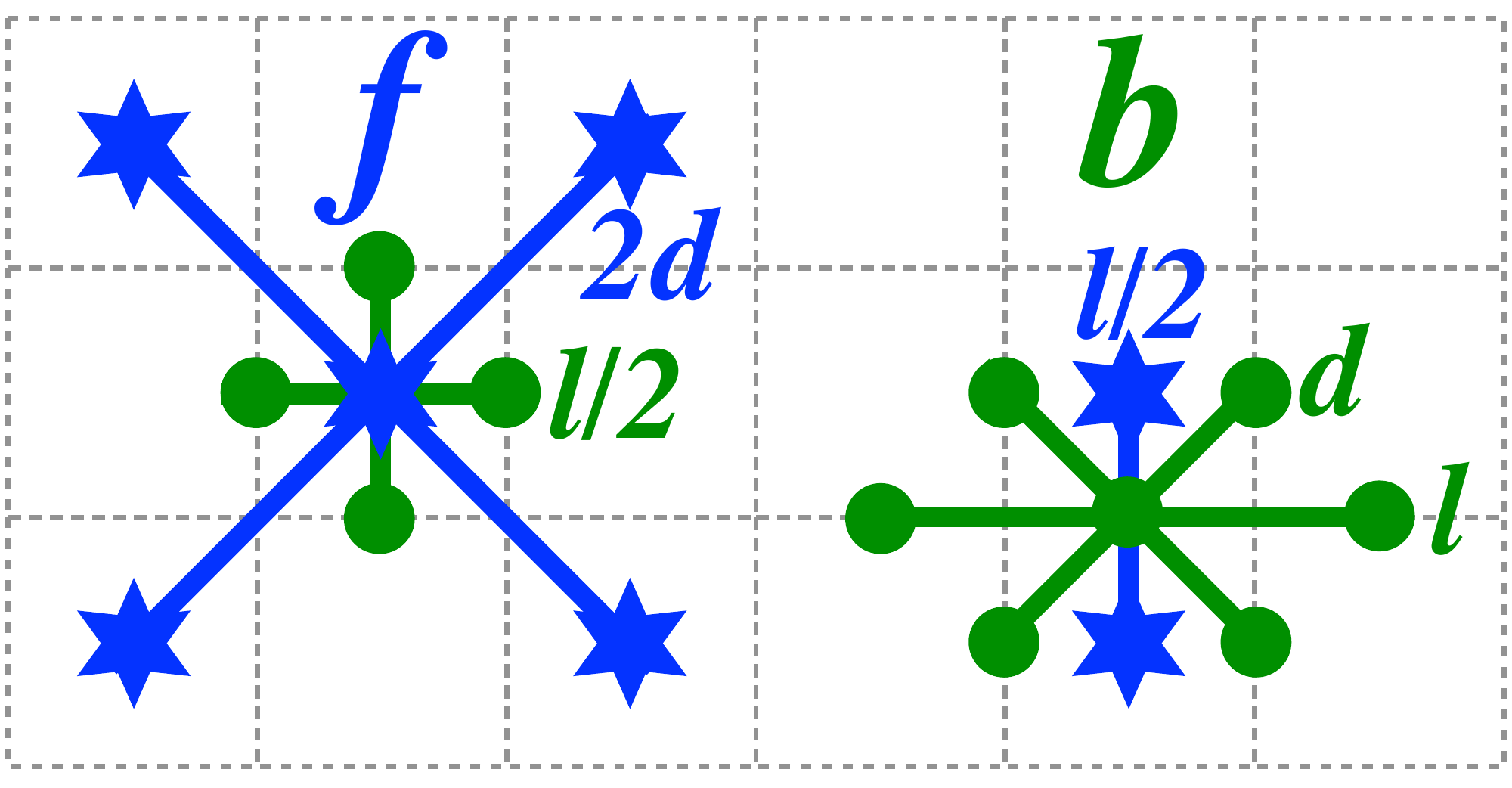}
\end{tabular}
\mbox{ \footnotesize or }
\begin{tabular}{c} 
  \includegraphics[height=1.1 cm]{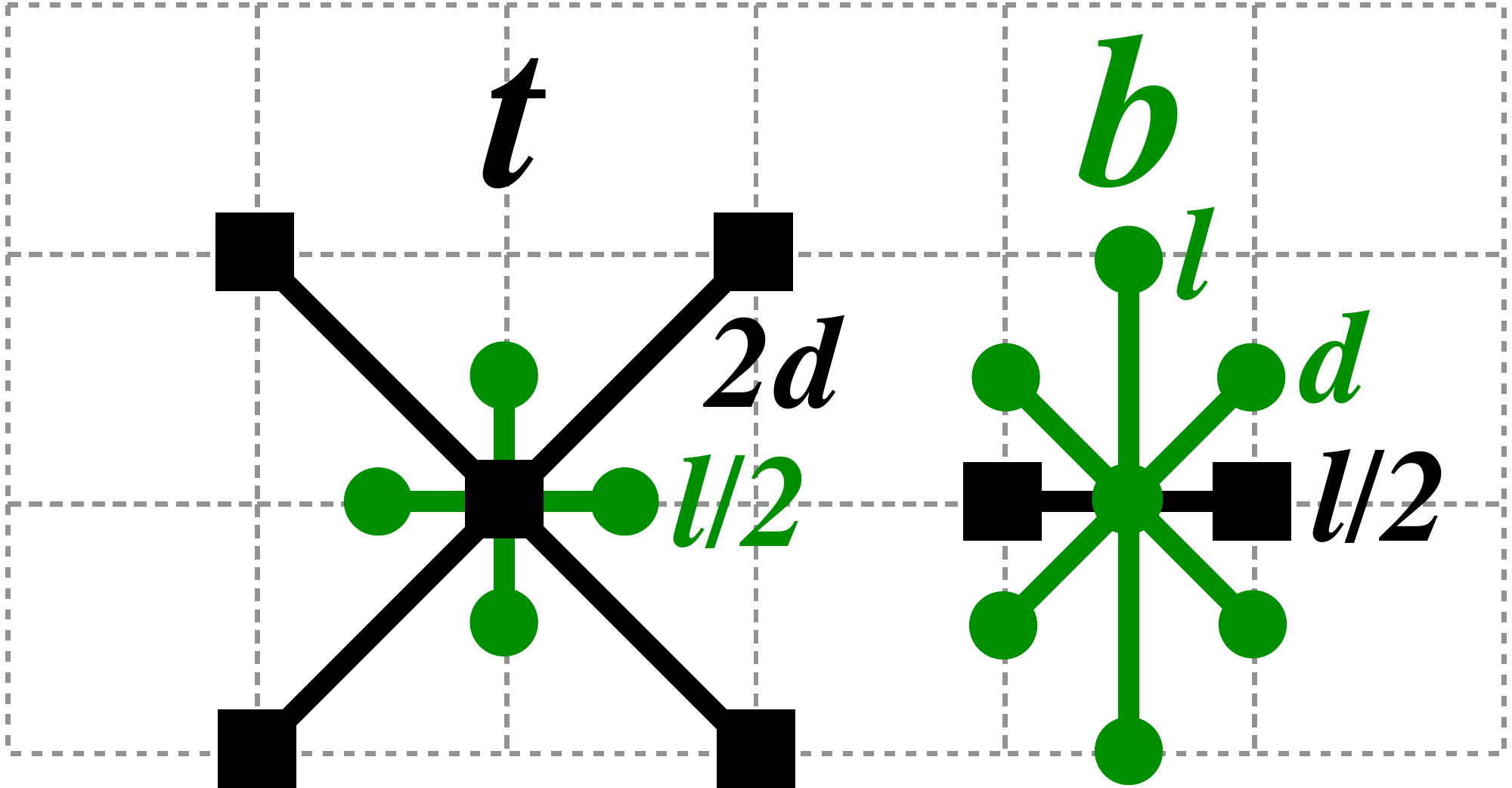}
\end{tabular}
      &
      \mbox{ \footnotesize $\frac{4}{2\alpha} \frac{2\nu_{ib}\nu_{bi}\left( \frac{l}{2} \right)^2+2\nu_{ib}\nu_{bb}^{d}d^2+\nu_{ib}\nu_{bb}^{l}l^2+\nu_{bi}\nu_{ii}\left( 2d \right)^2}{\nu_{bi}+2\nu_{ib}}$ }
      \\     
      
      \cline{2-4} 
      
      &      
\begin{tabular}{cccc} 
 \mbox{ \footnotesize $\nu_{fb}$ } &  \mbox{ \footnotesize $\nu_{tb}$ } &  \mbox{ \footnotesize $\nu_{bf}$ } &  \mbox{ \footnotesize $\nu_{bt}$ } \\
 \mbox{ \footnotesize $4$ } &  \mbox{ \footnotesize $4$ } &  \mbox{ \footnotesize $2$ } &  \mbox{ \footnotesize $2$ } \\
 \mbox{ \footnotesize $l/2$ } &  \mbox{ \footnotesize $l/2$ } &  \mbox{ \footnotesize $l/2$ } &  \mbox{ \footnotesize $l/2$ } \\
\end{tabular}
      &
\begin{tabular}{c}
 \vspace{-3 mm}\\
 \includegraphics[height=1.1 cm]{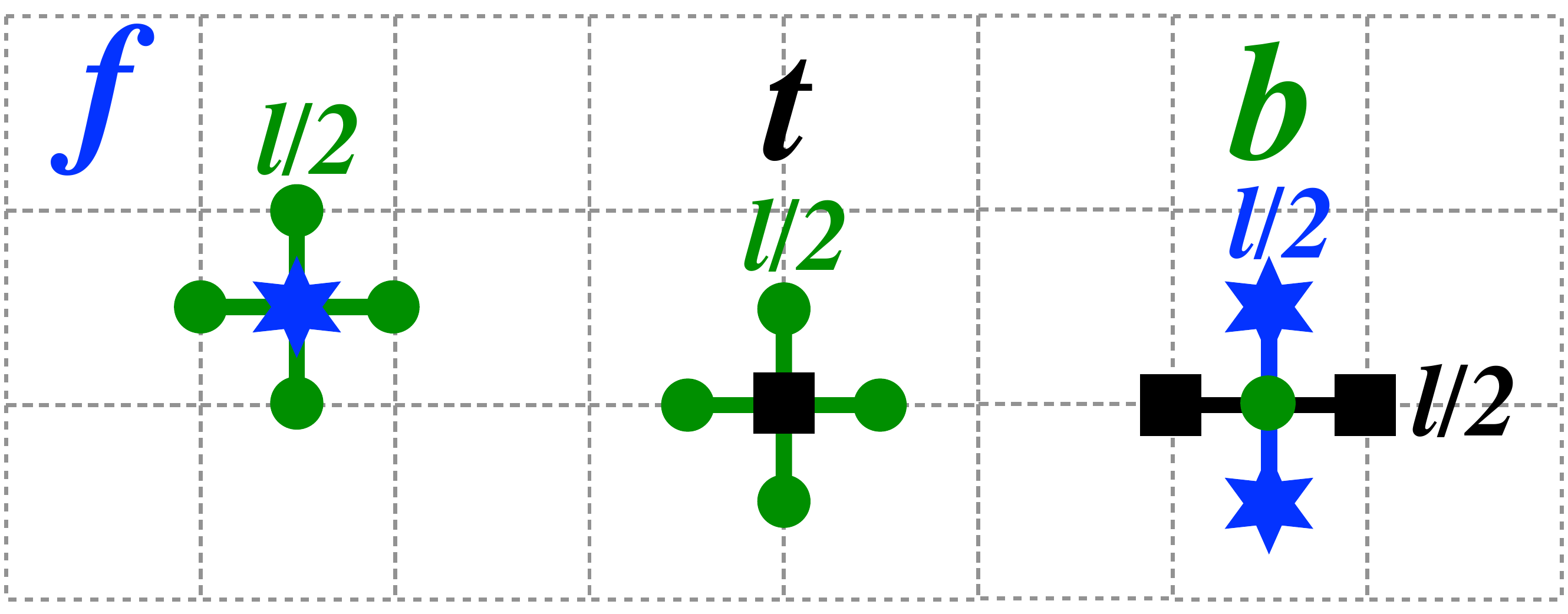}
\end{tabular}
      &
      \mbox{ \footnotesize $\frac{4}{2\alpha} \frac{2\nu_{fb}\nu_{tb}(\nu_{bf}+\nu_{bt})}{ 2\nu_{fb}\nu_{tb}+\nu_{tb}\nu_{bf}+\nu_{fb}\nu_{bt} }\left( \frac{l}{2} \right)^2$ }
      \\

      \hline
            
      \begin{sideways}\hspace{-7.5mm} \footnotesize Rectangular\end{sideways}      
     &
\begin{tabular}{cccccc} 
 \mbox{ \footnotesize $\nu_{ff}^{l}$ } &  \mbox{ \footnotesize $\nu_{ff}^{L}$ } &  \mbox{ \footnotesize $\nu_{ft}$ } &  \mbox{ \footnotesize $\nu_{tf}$ } &  \mbox{ \footnotesize $\nu_{tt}^{l}$ } &  \mbox{ \footnotesize $\nu_{tt}^{L}$ } \\
 \mbox{ \footnotesize $2$ } &  \mbox{ \footnotesize $2$ } &  \mbox{ \footnotesize $4$ } &  \mbox{ \footnotesize $4$ } &  \mbox{ \footnotesize $2$ } &  \mbox{ \footnotesize $2$ } \\
 \mbox{ \footnotesize $l$ } &  \mbox{ \footnotesize $L$ } &  \mbox{ \footnotesize $d$ } &  \mbox{ \footnotesize $d$ } &  \mbox{ \footnotesize $l$ } &  \mbox{ \footnotesize $L$ } \\
\end{tabular}
      &
\begin{tabular}{c} 
 \includegraphics[height=1.1 cm]{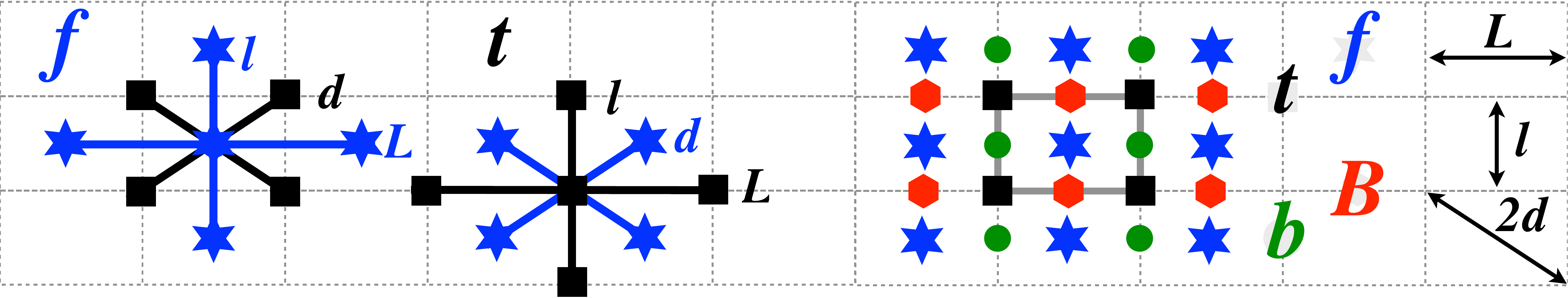} \\
\end{tabular}
      & 
      \begin{tabular}{c} 
      \\
      \mbox{ \footnotesize $\frac{2}{2\alpha} \frac{(\nu_{tf}\nu_{ff}^{l}+\nu_{ft}\nu_{tt}^{l})l^2+4\nu_{ft}\nu_{tf}d^2+(\nu_{tf}\nu_{ff}^{L}+\nu_{ft}\nu_{tt}^{l})L^2}{\nu_{ft}+\nu_{tf}}$ }
      \\
      \\
      \\
      \end{tabular}
      \\     

      \hline

      \begin{sideways}\hspace{-18mm}Triangular\end{sideways}      
      &
\begin{tabular}{r} 
\begin{tabular}{cccc} 
 \mbox{ \footnotesize $\nu_{ii}$ } &  \mbox{ \footnotesize $\nu_{ij}$ } &  \mbox{ \footnotesize $\nu_{ji}$ } &  \mbox{ \footnotesize $\nu_{jj}$ } \\
 \mbox{ \footnotesize $6$ } &  \mbox{ \footnotesize $3$ } &  \mbox{ \footnotesize $3$ } &  \mbox{ \footnotesize $6$ } \\
 \mbox{ \footnotesize $a$ } &  \mbox{ \footnotesize $l$ } &  \mbox{ \footnotesize $l$ } &  \mbox{ \footnotesize $a$ } \\
\end{tabular}
\\
\mbox{ \footnotesize $(i,j)=(f,h)$ or $(f,t)$ or $(h,t)$ }
\\
\end{tabular}
      &
\begin{tabular}{c} 
 \vspace{-3 mm}\\
 \includegraphics[height=2.1 cm]{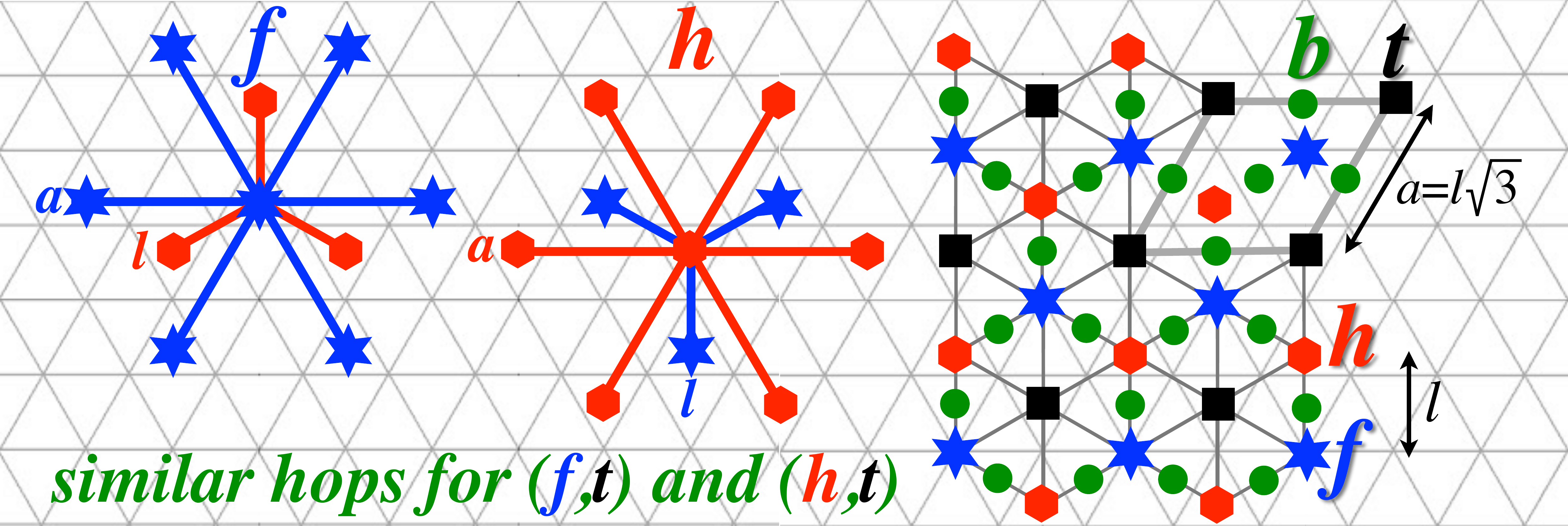}
\end{tabular}
      & 
      \mbox{ \footnotesize $\frac{6}{2\alpha} \frac{\nu_{ij}\nu_{ji}l^2+(\nu_{ij}\nu_{jj}+\nu_{ji}\nu_{ii})a^2}{\nu_{ij}+\nu_{ji}}$ } 
      \\     

      \cline{2-4} 

      &
\begin{tabular}{cccccc} 
 \mbox{ \footnotesize $\nu_{fb}$ } &  \mbox{ \footnotesize $\nu_{hb}$ } &  \mbox{ \footnotesize $\nu_{tb}$ } &  \mbox{ \footnotesize $\nu_{bf}$ } &  \mbox{ \footnotesize $\nu_{bh}$ } &  \mbox{ \footnotesize $\nu_{bt}$ } \\
 \mbox{ \footnotesize $3$ } &  \mbox{ \footnotesize $3$ } &  \mbox{ \footnotesize $6$ } &  \mbox{ \footnotesize $1$ } &  \mbox{ \footnotesize $1$ } &  \mbox{ \footnotesize $2$ } \\
 \mbox{ \footnotesize $l/2$ } &  \mbox{ \footnotesize $l/2$ } &  \mbox{ \footnotesize $a/2$ } &  \mbox{ \footnotesize $l/2$ } &  \mbox{ \footnotesize $l/2$ } &  \mbox{ \footnotesize $a/2$ } \\
\end{tabular}
      &
\begin{tabular}{c} 
 \vspace{-3 mm}\\
 \includegraphics[height=2.0 cm]{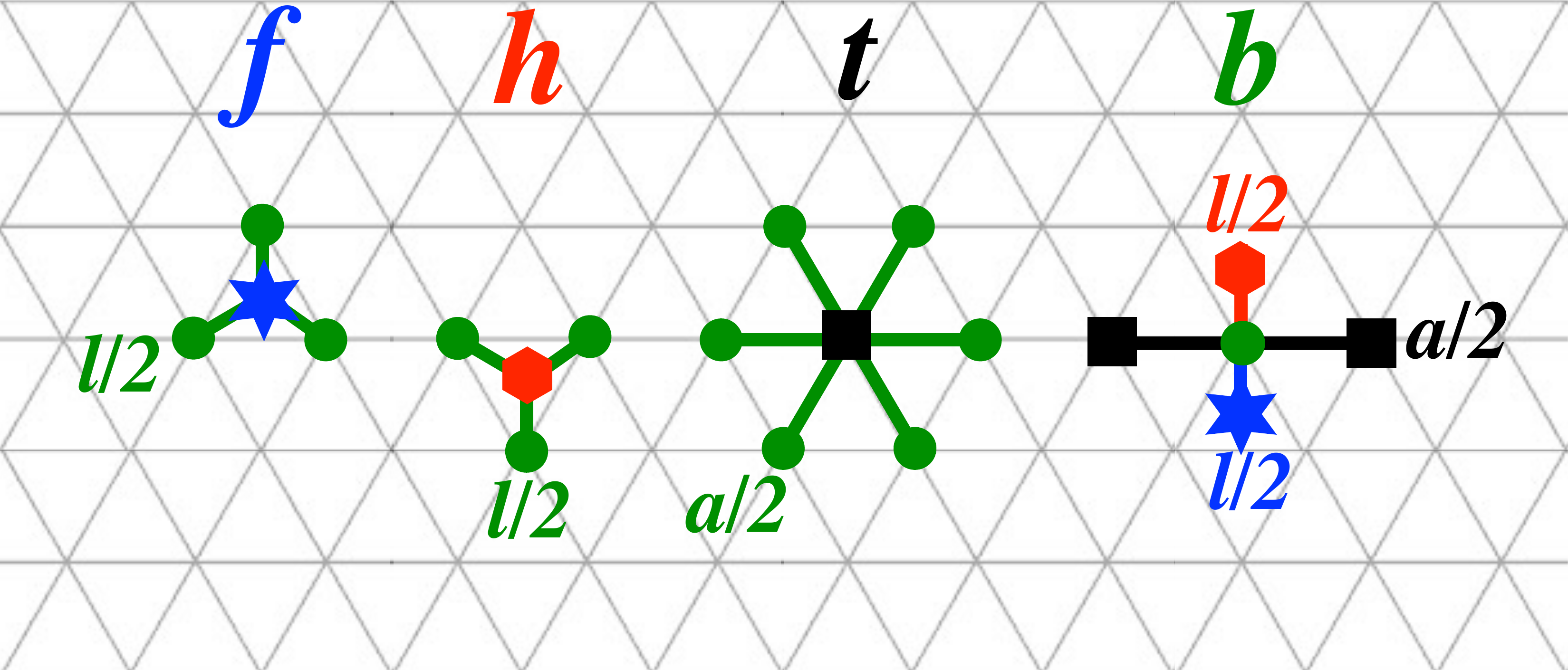}
\end{tabular}
      &
      \mbox{ \footnotesize $\frac{6}{2\alpha} \frac{\nu_{fb}\nu_{hb}\nu_{tb}(\nu_{bf}+\nu_{bh})\left( \frac{l}{2} \right)^2+2\nu_{fb}\nu_{hb}\nu_{tb}\nu_{bt}\left( \frac{a}{2} \right)^2}{\nu_{tb}(\nu_{hb}\nu_{bf}+\nu_{fb}\nu_{bh})+\nu_{fb}\nu_{hb}(\nu_{bt}+3\nu_{tb})}$ } 
     \\     

      \hline
      \hline
      
      \begin{sideways}\hspace{-20mm}Sequential hops\end{sideways}      
      &
\begin{tabular}{ccc} 
 \mbox{ \footnotesize $\nu_{12}$ } &  \mbox{ \footnotesize $\nu_{23}$ } &  \mbox{ \footnotesize $\nu_{31}$ } \\
 \mbox{ \footnotesize $n$ } &  \mbox{ \footnotesize $n$ } &  \mbox{ \footnotesize $n$ } \\
 \mbox{ \footnotesize $l$ } &  \mbox{ \footnotesize $l$ } &  \mbox{ \footnotesize $l$ } \\
\end{tabular}
      &
\begin{tabular}{c} 
 \vspace{-3 mm}\\
 \includegraphics[height=1.5 cm]{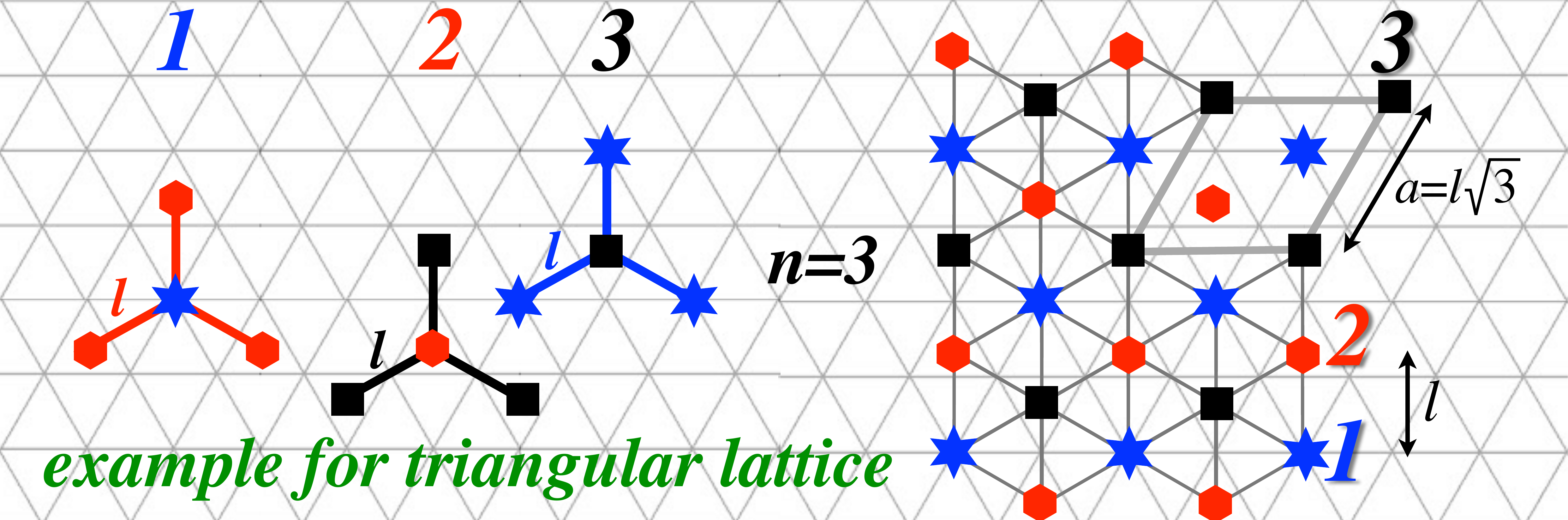}
\end{tabular}
      &
      \mbox{ \footnotesize $\frac{3n}{2\alpha} \frac{\nu_{12}\nu_{23}\nu_{31}}{\nu_{12}\nu_{23}+\nu_{23}\nu_{31}+\nu_{31}\nu_{13}} l^2$ } 
      \\     
      
      \cline{2-4} 

      &
\begin{tabular}{ccccc} 
 \mbox{ \footnotesize $\nu_{12}$ } &  \mbox{ \footnotesize $\nu_{23}$ } &  \mbox{ \footnotesize $\nu_{34}$ } &  \mbox{ \footnotesize $\nu_{41}$ } &  \mbox{ \footnotesize  }\\
 \mbox{ \footnotesize $n$ } &  \mbox{ \footnotesize $n$ } &  \mbox{ \footnotesize $n$ } &  \mbox{ \footnotesize $n$ } &  \mbox{ \footnotesize  }\\
 \mbox{ \footnotesize $d$ } &  \mbox{ \footnotesize $d$ } &  \mbox{ \footnotesize $d$ } &  \mbox{ \footnotesize $d$ } &  \mbox{ \footnotesize  }\\
\end{tabular}
      &
\begin{tabular}{c} 
 \vspace{-3 mm}\\
 \includegraphics[height=1.5 cm]{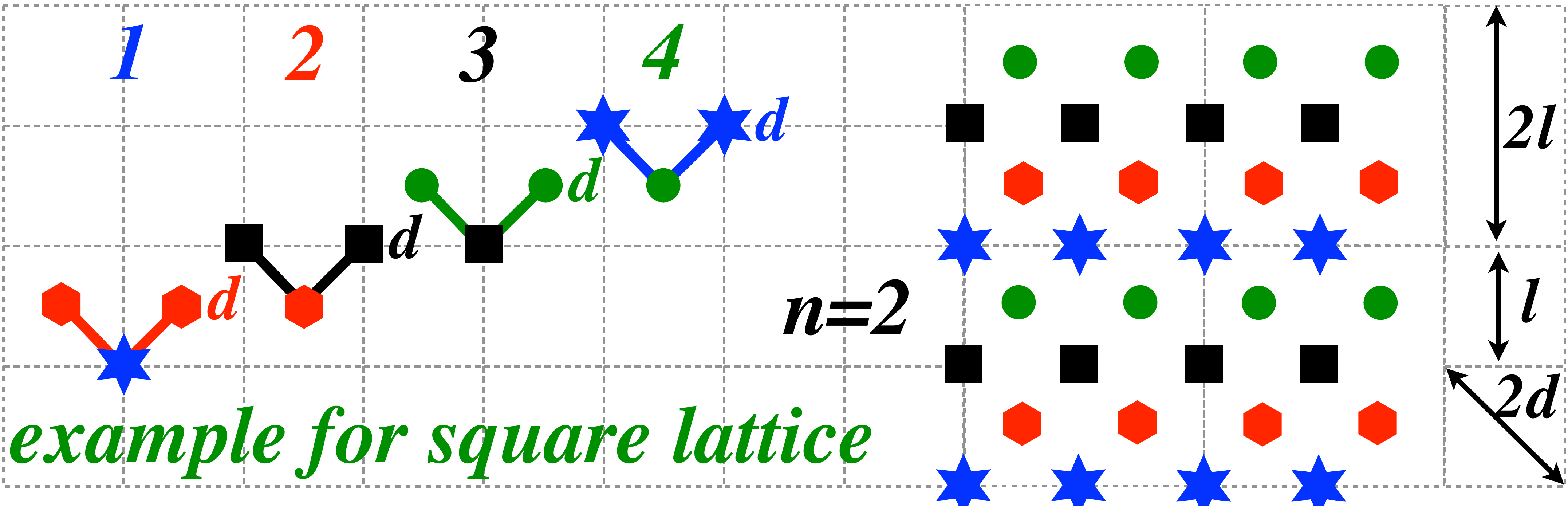} \\
\end{tabular}
      & 
      \mbox{ \footnotesize $\frac{4n}{2\alpha} \frac{\nu_{12}\nu_{23}\nu_{34}\nu_{41}}{\nu_{12}\nu_{23}\nu_{34}+\nu_{23}\nu_{34}\nu_{41}+\nu_{34}\nu_{41}\nu_{12}+\nu_{41}\nu_{12}\nu_{23}} d^2$ } 
      \\     
      
      \hline
      
   \end{tabular}

\end{adjustwidth}

\end{table*}

Table \ref{Table_ExampleEquations} shows a few example formulas obtained by applying equation \ref{D_5} and the outlined $\epsilon \rightarrow 0$ procedure for various systems, including triangular, rectangular and square lattices. More detailed tables containing all relevant hop combinations for each lattice are provided as {\bf Supplementary Data}. Although we restrict ourselves to the presentation of diffusivity expressions for 2D landscapes, equation \ref{D_5} is completely general and can be applied to 3D problems as well. In fact, the last two equations of Table \ref{Table_ExampleEquations} for three and four sequential jumps, respectively, are identical to those derived by Condit {\it et al.} \cite{Condit1969} and Birnie {\it et al.} \cite{Birnie1990}, respectively, for vacancy diffusion in three dimensions. Our procedure, however, is more general, taking into account any number of competing diffusion paths from any given site (parallel processes) in addition to any number of successive hops along any given diffusion path (sequential processes).

\section{Numerical validation}

\begin{figure*}[htb]
\begin{center}
\includegraphics[width=16.0 cm]{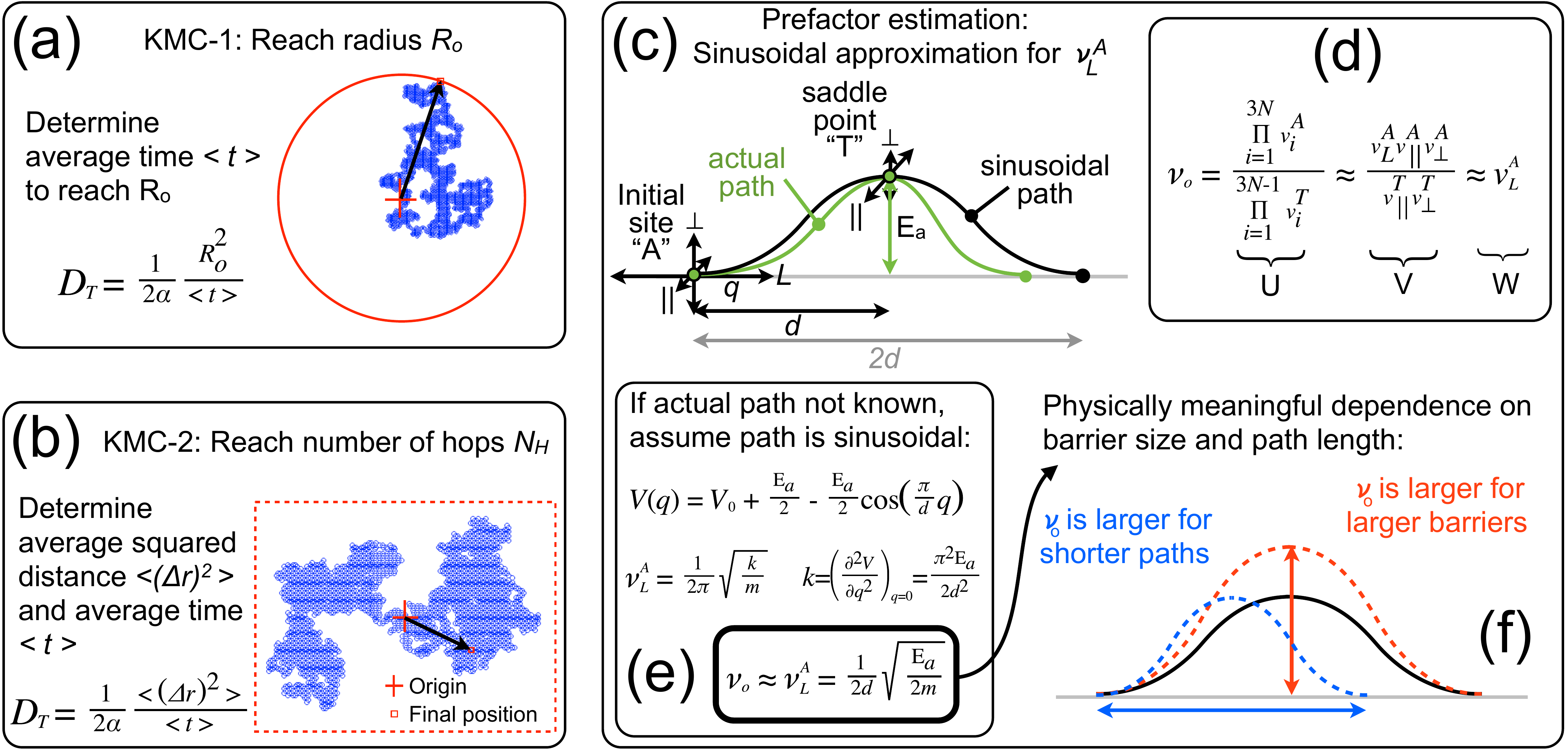}
\end{center}
\caption{
Illustration of the two simulation procedures used to perform KMC simulations of diffusion in this study: (a) KMC-1, (b) KMC-2. (c)-(e) Estimation of the attempt frequencies in this study. (f) Interpretation of the attempt-frequency dependence on the barrier size and path length.
}
\label{FigureKMC}
\end{figure*}

Based on the popularity of KMC simulations to determine tracer diffusivities \cite{Ala-Nissila, Naumovets, Gomer, Kellogg, Ferrando, Vattulainen1999, Salo2001, Lepage1998}, we now compare the values obtained from the previous formulas and those determined by KMC simulations. As described in  figure \ref{FigureKMC}(a)-(b), we perform two types of simulations. In the first type (KMC-1) the tracer is followed until it hits the perimeter of a circle of radius $R_{o} \gg l_{ij}$, repeating the process for $N_{RW}$ different random walks (RWs) in order to obtain an ensemble average of the time $< t >$ required to cover that distance, thus determining the diffusivity as $D_{T}^{\theta \approx 0}=\frac{1}{2\alpha}\frac{R_{o}^2}{< t >}$. In the second type of simulations (KMC-2) the tracer is followed until it performs a desired number of hops $N_{H}$, repeating the process for $N_{RW}$ different RWs to determine the average squared distance $< X^2 + Y^2 >$ covered by the tracer and the corresponding average time $< t >$, obtaining the diffusivity by using $D_{T}^{\theta \approx 0}=\frac{1}{2\alpha}\frac{ < X^2 + Y^2 > }{< t >}$.

Since the goal is to check the validity of the analytical expressions for the diffusivity, different hop rate values are used to probe situations where the rates have similar values / differ by several orders of magnitude. We also use realistic hop rates for several adsorbates, including Cu, Ag, Rb and Se on the Bi$_{2}$Se$_{3}$(0001) surface and in the vdW gap of this material. In this case the hop rates are expressed as $\nu=\nu_{0}e^{-E_{a}/k_{B}T}$, where the Boltzmann factor $e^{-E_{a}/k_{B}T}$ reflects the probability to perform the jump at temperature $T$ if the energy barrier is $E_{a}$, and $\nu_{0}$ is the attempt frequency, which reflects the dynamical coupling between the substrate phonons and the adparticle vibrations \cite{Ala-Nissila}. The actual values for these hop rates are obtained by determining the energy barriers through labor-intensive DFT calculations (see the {\bf Supplementary Data}) and estimating the attempt frequencies by the method described in  figure \ref{FigureKMC}(c)-(e), leading to:
\begin{equation}
\nu_{o} \approx \nu_{L}^{A} \approx \frac{1}{2d}\sqrt{ \frac{E_{a}}{2m} }.
\label{Approximation_for_nu0} 
\end{equation}
Here, $E_{a}$ is the energy barrier for the hop, $d$ is the separation between the initial site $A$ and the saddle point $T$ (transition state) and $m$ is the adatom mass.

This estimate can be considered as an alternative to (i) the typical assumption of equal prefactors for all hop rates \cite{Ala-Nissila, Ratsch1998, Yildirin2006, Yildirin2007, Lepage1998}, and (ii) the large computational cost to determine all the vibrational mode frequencies $\nu_{i}^{A}$ and $\nu_{i}^{T}$ for the adatom and substrate at the initial and saddle configurations by using DFT methods \cite{Ala-Nissila, Ratsch1998, Yildirin2006, Yildirin2007} (see expression 'U' for the attempt frequency in  figure \ref{FigureKMC}(d)). Due to typical cancellations of the vibrational modes of the substrate \cite{Yildirin2007}, the prefactor $\nu_{o}$ is usually approximated by the Vineyard equation (expression 'V' in  figure \ref{FigureKMC}(d)). Owing to compensation effects between the surface-parallel ($\nu_{\parallel}^{A}$ and $\nu_{\parallel}^{T}$) and surface-normal ($\nu_{\perp}^{A}$ and $\nu_{\perp}^{T}$) vibrational frequencies of the diffusing atom \cite{Yildirin2007}, the Vineyard formula is approximated by just keeping the frequency of the longitudinal vibrations (along the diffusion path) of the atom at the initial site ($\nu_{L}^{A}$, see expression 'W' in  figure \ref{FigureKMC}(d)). Our estimate consists on approximating the longitudinal path by a sinusoidal path, resulting in a simple and physically meaningful expression for $\nu_{L}^{A}$ in terms of the energy barrier, hop distance and adatom mass, as described in  figure \ref{FigureKMC}(f) and contained in equation \ref{Approximation_for_nu0}. Although the actual energy path can be asymmetric with respect to the saddle point T, the part after this point is irrelevant for the rate calculation and in our approximation it is considered to be a reflection of the part before it.

For all considered systems the simulated and calculated diffusivities agree extremely well with each other (see Table S5 of the {\bf Supplementary Data}), thus concluding that the proposed formulas are suitable to discuss the low coverage tracer diffusivity of typical diffusion species in complex energy landscapes without any need to perform the corresponding KMC simulations.

\section{Application to topological insulators}

\begin{figure}[htb]
\begin{center}
\includegraphics[width=11.0 cm]{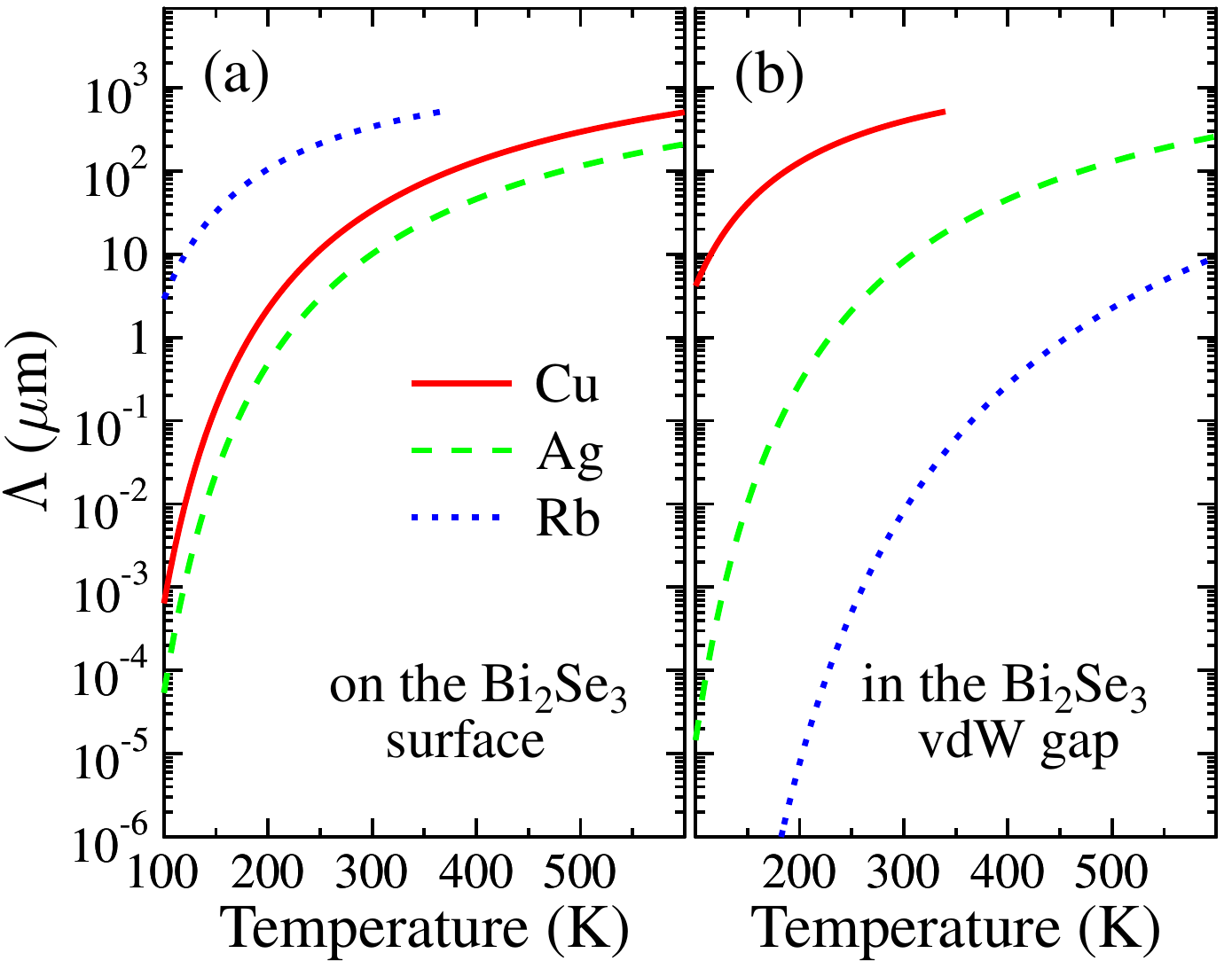}
\end{center}
\caption{
Calculated diffusion length $\Lambda$ (in logarithmic scale) as a 
function of temperature for Cu, Ag and Rb adatoms: (a) 
on the Bi$_{2}$Se$_{3}$(0001) surface, and (b) in the 
Bi$_{2}$Se$_{3}$ vdW gap. Diffusion time: 1 min. 
}
\label{Figure_DiffusionLength_vs_Temp}
\end{figure}

Encouraged by the validation of the diffusivity formulas we now consider the temperature dependence of the diffusion of Cu, Ag and Rb adatoms on the Bi$_{2}$Se$_{3}$(0001) surface and the corresponding intercalation of these adsorbates in the Bi$_{2}$Se$_{3}$ vdW gap.
Figure \ref{Figure_DiffusionLength_vs_Temp}(a)-(b) provides 
the calculated diffusion length $\Lambda=\sqrt{2\alpha D^{\theta 
\approx 0}_{T} t}$ as a function of temperature 
for the three considered adatoms when they diffuse on the 
surface and in the vdW gap, respectively. 
The evaluations have been done on the basis of the formulas 
obtained for $D^{\theta \approx 0}_{T}$ (see fifth column 
of Table S5 of the {\bf Supplementary Data}). 
For diffusion with a single barrier $E_{a}$, the underlying assumption that the hops can be treated as rare events (as compared to the fast vibrations around the adsorption sites) is valid if $E_{a} > 4k_{B}T$ (see Ref. \cite{Ala-Nissila}). Correspondingly, each displayed curve in  figure \ref{Figure_DiffusionLength_vs_Temp} terminates at $T_{\rm max}=E_{a, {\rm max}}/4k_{B}$, where $E_{a,\rm max}$ is the maximum barrier experienced by the corresponding adatom. As an example, the diffusion of Rb on the surface experiences two barriers: $E_{fh}$ = 0.127 eV and $E_{hf}$ = 0.104 eV. Thus, $E_{a,{\rm max}}$ = 0.127 eV and $T_{\rm max}$ = 368 K. Similarly, for Cu in the vdW gap we have $T_{\rm max}$ = 342 K. Nevertheless, the total temperature range is restricted to 600 K since the desorption of Cu is reported to start at $\sim$550 K \cite{Ikeda.slr1996}.

It can be seen from figure \ref{Figure_DiffusionLength_vs_Temp} 
that the hierarchy of the diffusion length is 
$\Lambda_{\rm Rb} > \Lambda_{\rm Cu} > \Lambda_{\rm Ag}$ on the surface and 
$\Lambda_{\rm Cu} > \Lambda_{\rm Ag} > \Lambda_{\rm Rb}$ in the vdW gap. 
The Rb (Cu) atoms are the most mobile species on the surface (in the vdW gap), 
capable of covering more than 1~$\mu$m within one minute even at 100~K. 
However, the vdW (surface) diffusion length of the Rb (Cu) atoms is 
much lower than that on the surface (in the vdW), which is due
to significantly higher diffusion barriers. Interestingly, 
the Ag atoms travel with almost equal rates on the surface and 
in the vdW gap. 
\begin{figure}[htb]
\begin{center}
\includegraphics[width=11.0 cm]{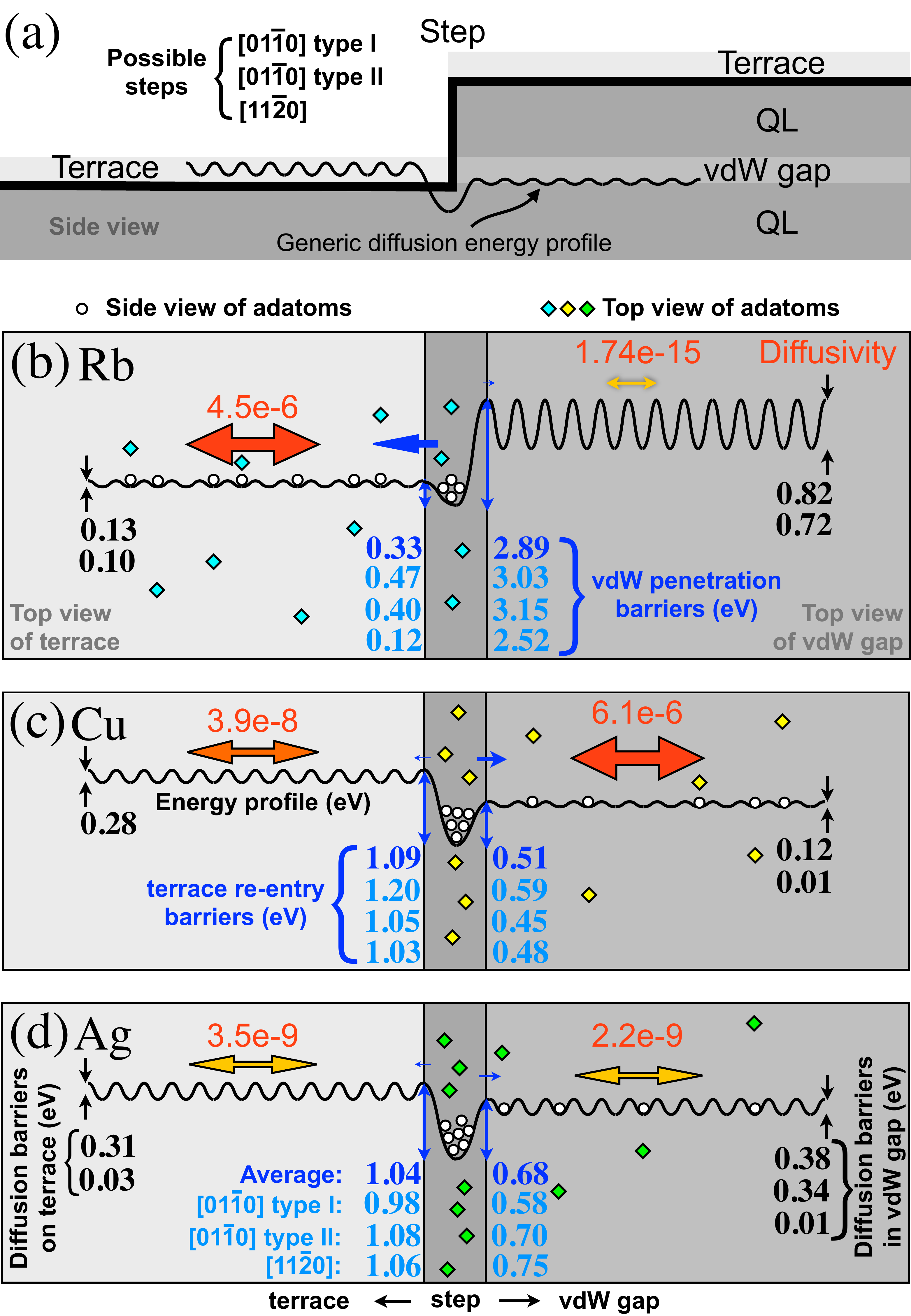}
\end{center}
\caption{
(a) Geometrical alignment between terrace and vdW gap at a morphological step. 
(b)-(d) Schematic representation of the low coverage diffusivity (in cm$^2$s$^{-1}$) 
of Rb, Cu and Ag on the terrace and in the 
vdW gap of Bi$_{2}$Se$_{3}$ at room temperature. Larger arrows denote
larger diffusivities. Our DFT-calculated diffusion barriers (in eV) are shown in boldface for: (left) terrace diffusion, (center-left) terrace re-entry, (center-right) vdW-gap penetration and (right) vdW-gap diffusion.
}
\label{TerraceStepVdW}
\end{figure}

Let us now bring the discussion closer to the available experiments 
\cite{Bianchi2, Wang, Ye.arxiv2011}. These indicate indirectly the occurrence of
partial \cite{Bianchi2} or almost complete \cite{Wang, Ye.arxiv2011}   
intercalation of the metal adatoms inside the Bi$_2$Se$_3$ vdW gap
at room and higher temperatures. Recently, it has been argued 
\cite{Otrokov.jetpl2012} that the intercalation of the metal atoms in 
the Bi$_2$Se$_3$ vdW gap is step-mediated, in the sense that they 
penetrate into the vdW gap after reaching the steps, which are typically
present at the Bi$_2$Se$_3$ surface \cite{Wang}. 
This is favored by the geometrical alignment between the terrace and the vdW gap in a stepped surface, 
as schematically shown in  \ref{TerraceStepVdW}(a).
At the same time, 
penetration of Cu \cite{Wang} and Ag \cite{Otrokov.jetpl2012} in the 
vdW gap via interstitials and/or vacancies of the topmost Bi$_2$Se$_3$ 
QL is significantly less probable 
due to high energy barriers. 
Therefore, one may also rule out the possibility of {\em vertical} 
penetration of the Rb atom, whose covalent radius is 1.5 (1.66) times larger than 
that of Ag (Cu), whereupon the diffusion barriers are expected to 
be even higher.

By using large and small double-head arrows, figure \ref{TerraceStepVdW}(b)-(d) presents a graphical description of the relative diffusivity of the three types of adatoms on the terraces and within the vdW gaps. In addition, the figure also provides the relative rates to enter the vdW gap and to bounce back to the terrace by assigning them large/small unidirectional arrows. Moreover, the figure collects all the diffusion barriers determined by our DFT calculations for the three adatoms on the terrace and in the vdW gap, as well as for their penetration into the vdW gap and re-entry into the terrace for two different step orientations, namely, [$11\bar20$] and [$01\bar10$], the latter having two possible atomic terminations \cite{Otrokov.jetpl2012}. To ease the discussion, the three barriers for vdW penetration (terrace re-entry) for each adatom are algebraically averaged and used to estimate the vdW penetration (terrace re-entry) rate of that atom, accordingly assigning them large/small unidirectional arrows. We center the discussion at room temperature (295 K) and long diffusion times.

The Rb atoms have the largest terrace and smallest vdW diffusivities, 
with a very low vdW-penetration rate and a rather large terrace-reentry rate.
Thus, the Rb atoms are expected to quickly diffuse across 
the terraces and hit the steps, where a small fraction will remain 
trapped while the majority will bounce back to the terraces. 
This is described schematically in figure \ref{TerraceStepVdW}(b) by 
drawing a large number of Rb atoms in the terrace region, with 
additional atoms at the step, while hardly any atoms in the vdW gap. 
In comparison, the Cu atoms are characterized by moderate terrace 
and largest vdW diffusivities, with the largest vdW-penetration rate 
and a low terrace-reentry rate. Accordingly, the Cu atoms need a 
longer time to arrive at the steps but they eventually penetrate 
with relative ease into the vdW gap, where they diffuse rather fast. 
Thus, the Cu atoms are expected to mostly intercalate in the vdW 
gap, although a notable fraction will remain at the steps, as sketched in
figure \ref{TerraceStepVdW}(c). 
Finally, the Ag atoms display slightly lower terrace diffusivity 
as compared to Cu and a medium diffusivity in the vdW gap 
among the three species under consideration.
Having reached the steps, the Ag atoms are forced to linger along them 
due to the large terrace-reentry barrier and significant vdW-penetration 
barrier (figure \ref{TerraceStepVdW}(d)). Since the latter is smaller, 
the Ag atoms are expected to gradually intercalate into the vdW gap.
The behavior is similar to that of Cu, although intercalation is slower for Ag. 
Valid only for the low coverage limit, these trends are in good qualitative agreement with those 
from available experiments \cite{Bianchi2, Wang, Ye.arxiv2011}.

\section{Conclusion}

Focusing on the analysis of adsorbate diffusion on surfaces and within the two-dimensional gap of layered materials, we present a combination of formulas for future reference and their application to intercalation in a topological insulator (Bi$_{2}$Se$_{3}$). We start by presenting a general expression to determine the average motion of the diffusing particles at low densities in complex, periodic energy landscapes consisting of various energy barriers located between distinct adsorption sites in any number of dimensions. For adsorbate diffusion in two dimensions, formulas are provided for the low coverage tracer diffusivity for complex combinations of hop rates in systems with triangular, rectangular and square symmetry. The analytical expressions are validated against Kinetic Monte Carlo simulations, obtaining an excellent agreement between the calculated and simulated diffusivities. Thus, one can determine the overall diffusion coefficient without performing the KMC simulations. Based on diffusion rates from energy barriers obtained by labor-intensive density functional theory calculations, we analyze the temperature dependence of the diffusion of Cu, Ag and Rb on the Bi$_{2}$Se$_{3}$(0001) surface and within the van der Waals (vdW) gap inside the (layered) crystal. We also analyze the occurrence of adsorbate intercalation due to the alignment between the vdW gaps and terraces on [11$\overline2$0]- and two types of [01$\overline1$0]-stepped surfaces of this topological insulator. At room temperature and low coverage, we conclude that the Rb atoms quickly diffuse across the terraces and hit the steps, where a small number remain trapped while the rest bounce back into the terraces. Thus, Rb is expected to partially decorate the steps while remaining present on the terraces. In comparison, the Cu atoms take a longer time to arrive to the steps, eventually penetrating with relative ease into the vdW gap. Thus, Cu atoms are expected to mostly intercalate into the vdW gap while partially remaining at the steps. Ag atoms are expected to take an even longer time to diffuse across the terraces, eventually penetrating into the vdW gap with time but meanwhile remaining around the steps.


\section*{Acknowledgments}
We acknowledge support by the Ram\'on y Cajal Fellowship Program by the Spanish Ministry of Science and Innovation, the JAE-Doc grant from the 'Junta para la Ampliaci\'on de Estudios' -program co-funded by FSE, the University of the Basque Country (Grant No. GIC07IT36607), the Spanish Ministry of Science and Innovation (Grant No. FIS2010-19609-C02-00), the Basque Government through the NANOMATERIALS project (Grant  IE05-151) under the ETORTEK Program (iNanogune) and the Ministry of Education and Science of Russian Federation (state task No. 2.8575.2013). The DFT calculations were performed on the SKIF-Cyberia supercomputer of Tomsk State University as well as in Donostia International Physics Center.

\section*{References}

\end{document}